\documentclass[conference]{IEEEtran}
\usepackage{cite}
\usepackage{amsmath,amssymb,amsfonts}
\usepackage{algorithmic}
\usepackage{graphicx}
\usepackage{textcomp}
\usepackage{xcolor}

\def\BibTeX{{\rm B\kern-.05em{\sc i\kern-.025em b}\kern-.08em
		T\kern-.1667em\lower.7ex\hbox{E}\kern-.125emX}}

\newcommand\blankfootnote[1]{%
	\let\thefootnote\relax\footnotetext{#1}%
	\let\thefootnote\svthefootnote%
}

\begin{document}
	
	\title{Working in Harmony: Towards Integrating RSEs into Multi-Disciplinary CSE Teams}
	
	\author{\IEEEauthorblockN{Miranda Mundt}
		\IEEEauthorblockA{\textit{Department of Software Engineering and Research} \\
			\textit{Sandia National Laboratories}\\
			Albuquerque, NM \\
			mmundt@sandia.gov}
		\and
		\IEEEauthorblockN{Reed Milewicz}
		\IEEEauthorblockA{\textit{Department of Software Engineering and Research} \\
			\textit{Sandia National Laboratories}\\
			Albuquerque, NM \\
			rmilewi@sandia.gov}}
	
	\maketitle
	
	\begin{abstract}
		Within the rapidly diversifying field of computational science and engineering (CSE), research software engineers (RSEs) represent a shift towards the adoption of mainstream software engineering tools and practices into scientific software development. An unresolved challenge is the need to effectively integrate RSEs and their expertise into multi-disciplinary scientific software teams. There has been a long-standing ``chasm''  between the domains of CSE and software engineering, and the emergence of RSEs as a professional identity within CSE presents an opportunity to finally bridge that divide. For this reason, we argue there is an urgent need for systematic investigation into multi-disciplinary teaming strategies which could promote a more productive relationship between the two fields.

	\end{abstract}
	
	\blankfootnote{Presented at the Workshop on the Science of Scientific-Software Development and Use, sponsored by U.S. Department of Energy, Office of Advanced Scientific Computing Research, Dec 13–15, 2021.
		
		Sandia National Laboratories is a multimission laboratory managed and operated by National Technology \& Engineering Solutions of Sandia, LLC, a wholly owned subsidiary of Honeywell International Inc., for the U.S. Department of Energy’s National Nuclear Security Administration under contract DENA0003525. SAND2021-14806 C.}
	
	
	\section{Challenge}\label{challenge}
	
	In the past decade, the computational science and engineering (CSE) community has grown to include more software engineering (SE) practitioners and researchers. This is most visibly evident on the practice front with the emergence of the research software engineer (RSE) as a professional designation~\cite{baxter2012}. These RSEs bring with them valuable expertise in the science and engineering of software. Software engineering is the ``disciplined application of engineering, scientific, and mathematical principles and methods to the economical production of quality software''~\cite{humphrey1988software}. Every engineering discipline has counterparts in science, and for software engineering, that is software science (also known as software engineering research): the systematic study of software systems and their development, operation, and maintenance~\cite{shen1983software}. These twin fields, which first emerged as a response to the ``software crisis'' of the late 1960s, are united in the pursuit of evidence-based best practices for the development of software. 
	
	In this effort, the SE community enjoyed considerable success: many of the tools, techniques, and processes that are employed by developers today are the product of a diligent synthesis of research and practice. We believe that similar progress can be made in scientific software development~\cite{milewicz2020towards}. What is not yet clear, however, is how to integrate SE expertise into multi-disciplinary computational science teams. Situating RSEs inside projects does not, by itself, guarantee effective teaming or the adoption of SE best practices; if we truly mean to include and enfranchise RSEs within the CSE community, we must address this challenge. 
	
	There has long been a divide -- which notable commentators have described as a ``chasm''~\cite{kelly2007software,faulk2009scientific,storer2017bridging} -- between CSE and software engineering. CSE researcher-developers have historically maintained an independent community of practice outside of mainstream software engineering\cite{hannay2009scientists} -- one with its own terms, techniques, values, and norms. We have seen how these differences can spur conflict: researchers feeling that software engineers fail to appreciate the domain-specific nuances of scientific computing, and software engineers feeling that researchers only value quality and craftsmanship insofar as it advances the science.
	
	This is a common problem in multi-disciplinary team contexts. To quote one scholar on multi-disciplinary teaming, ``a multi-disciplinary team without differences is a contradiction in terms''~\cite{ovretveit1995team}. Within multi-disciplinary software development specifically, Burnell et al. has noted how culture differences, conflicts in working patterns, and a lack of understanding of other disciplines can impede communication and effective teaming~\cite{burnell2002teaching}. While it is imperative that the CSE community adopt and adapt SE best practices to meet the ever-growing demand for scientific computing, this can only succeed if we resolve those communication barriers. There is an urgent need for systematic research in this space to demonstrate effective teaming strategies that can bridge the divide.
	
	\section{Why now?}
	
	As we mentioned previously, research institutions in recent years have moved to bring RSEs into the fold. The creation of both RSE groups at national labs and universities (\textit{e.g.,} ~\cite{willenbring2020moving,milewicz2020research}) as well as national associations promoting the interests of those RSEs~\cite{ukRSEWebsite,usSEWebsite,deRSEWebsite,nlRSEWebsite} have been promising developments. Whereas before SE was viewed as something separate from and outside of CSE, we can now point to RSEs working on the frontlines of scientific software development. Having researchers and RSEs in the same room has been a necessary first step towards reconciliation.
	
	Likewise, we have seen a revaluation of SE concepts and practices in science and mathematics. As noted in a 2018 SIAM report by Rude et al., software itself is the foundation for collaboration and discovery in CSE, and this makes software engineering ``central to any effort to increase CSE software productivity''~\cite{rude2018research}. According to the report, not only must scientists and mathematicians become more conversant in software engineering concepts and techniques, CSE education and training should emphasize cross-disciplinary communication and collaboration so that researchers can work more effectively with software engineers and others. 
	
	Software has become an invaluable tool in the pursuit of scientific research, and we are finally at a stage where we can have a productive dialogue on the roles of software engineering and software engineers in the development of that software.

	\section{Opportunity}
	
	Incorporating SE into CSE through better teaming is, at its core, a research challenge. How do we create robust research teams that leverage RSEs and their expertise? How should professionals from different computing backgrounds interact? How do we evaluate the quality of their teaming? Can we provide empirical evidence for the effectiveness of certain multi-disciplinary teaming strategies?  There is a clear need for experience reports and case studies addressing these questions to provide an evidentiary basis for further meta-analyses. Within the Department of Software Engineering and Research at Sandia National Laboratories, we have found that embedding RSEs into research efforts has resulted in better teaming, productivity, and quality in software\cite{mundt2020overcoming}, and this dovetails with similar reports elsewhere (\textit{e.g.,} \cite{cohen2020four}). 
	
	At the present, a chief concern for the authors is the lack of a well-defined framework for recreating these successes. Reproducibility is the foundation of all science, and validated models for how multi-disciplinary scientific software teams can work together would be integral to new teams replicating the successes of others. Research in this space could provide a better understanding of team composition and consensus-building strategies between researchers and RSEs. This would build upon the existing body of scholarship into cross-domain collaborative software development teams, such as in the realm of video game development~\cite{freeman2019exploring}, and we hope to see similar results in the CSE domain.
	
	Moreover, we believe that addressing this RSE teaming challenge supports the broader investigation into the science of scientific software development. There is a panoply of as-yet unresolved questions in software science that have been held back by limited access to scientific software teams. Maintaining a strong SE presence within CSE will facilitate software engineering research into tools, techniques, and methodologies.

\bibliographystyle{IEEEtran}
\bibliography{ascr_workshop.bib}

\end{document}